\documentclass[prc,twocolumn]{revtex4}

\usepackage{graphicx}
\usepackage{dcolumn}
\usepackage{bm}
\usepackage{amsmath}

\begin{document}

\title{Microscopic calculations of weak decays in superheavy nuclei}
\author{P.~Sarriguren}
\email{p.sarriguren@csic.es}

\affiliation{Instituto de Estructura de la Materia, IEM-CSIC, Serrano 123, 
E-28006 Madrid, Spain}
\date{\today}

\begin{abstract}
Half-lives of  $\beta^+$ decay and electron capture are studied in some selected
superheavy nuclei produced in hot-fusion reactions, namely, $^{290}$Fl, $^{293}$Mc, 
$^{294}$Lv, and $^{295}$Ts. The nuclear structure is described microscopically from 
deformed self-consistent Skyrme Hartree-Fock mean-field calculations that include 
pairing correlations. The sensitivity of the half-lives to deformation and to the 
$Q_{EC}$ energies, which are still not determined experimentally, are studied. The 
results are compared with phenomenological $\alpha$-decay half-lives, showing 
that the latter decay mode is dominant in this mass region.

\end{abstract}

\maketitle


\section{Introduction}

The search for new regions of nuclear stability in superheavy nuclei (SHN) is a very 
active and successful line of research that has already led to the discovery of a 
large number of new elements \cite{hofmann_00,oganessian_04,oganessian_07,hamilton_13,oganessian_15,hong_17,giuliani_19,hofmann_16}.
First calculations of binding energies within macroscopic-microscopic models 
\cite{hofmann_16,myers_66,sobiczewski_66,nilsson_68,patyk_91,moller_94,smolanczuk_95}
predicted the existence of "islands of stability" for spherical SHN $Z=114$ and 
$N=184$, as well as for deformed nuclei with  $Z=108$ and  $N=162$. In these models 
a macroscopic term, usually derived from a deformed liquid-drop model, is 
complemented with a microscopic part that includes a shell correction derived from 
a shell model calculation. Purely microscopic calculations showed that the location 
of the shell closures in SHN is not very robust, but model dependent. Different 
selfconsistent relativistic and non-relativistic mean-field models 
\cite{rutz_97,kruppa_00,bender_01,bender_03,meng_06,dobaczewski_15} predict closure 
of spherical shells at $Z=114,\ N=184$, $Z=120,\ N=172$, and $Z=126,\ N=184$,
depending on the interactions and their parametrizations. Note that the 
macroscopic-microscopic calculations performed with the modified two-center 
shell model \cite{kuzmina_12} reveal quite strong shell effects at $Z=120-126$ and 
$N=184$, in agreement with the self-consistent mean-field treatments.

In parallel, different experimental strategies were successfully carried out to 
reach the theoretically predicted "islands of stability" for SHN. The cold-fusion 
approach was used to synthesize SHN with $Z=107-112$  in reactions with target magic 
nuclei ($^{208}$Pb and  $^{209}$Bi) and massive projectiles, such as $^{50}$Ti, 
$^{54}$Cr, $^{58}$Fe, $^{62,64}$Ni, and $^{70}$Zn \cite{hamilton_13}. These reactions 
are cold in the sense that the compound nucleus has low excitation energy and only 
one or two neutrons evaporate. However, the method is not applicable for reaching 
heavier nuclei, in particular  around $Z=114,\ N=184$, because of the fast 
decrease of the production cross sections for increasing charge of the projectile.
To overcome this difficulty, a second strategy was developed, using more asymmetric 
reactions (less Coulomb repulsion) with both target and projectile having a large 
neutron excess. Following this strategy, long-lived actinide nuclei from $^{238}$U 
to$^{249}$Cf were used as targets,  whereas the double magic nucleus $^{48}$Ca was 
used as a beam. These so called  hot-fusion reactions result in the production of 
SHN with $Z$=112--118 in the neutron-evaporation channels ($xn$-channels)
\cite{oganessian_15}. The main advantage of these reactions is that  the Coulomb 
force becomes weaker as compared to the cold-fusion reactions and the probability 
of forming a compound nucleus increases dramatically. However, in hot-fusion 
reactions the compound nucleus formed is highly excited and more neutrons are 
evaporated. After evaporation of $x$ neutrons ($x=2-5$), nearly all the new nuclei 
produced in the hot-fusion reactions undergo a chain of $\alpha$ decays ending 
with a spontaneous fission. Identification of the associated $\alpha$-decay 
chains is the link to establish the original SHN.

Further experimental extension of the SHN region in the direction of the magic 
neutron number $N=184$,  where the center of the "island of stability" is predicted, 
is limited in the  $xn$-channels by the number of available stable projectiles and 
targets and the small production cross sections. One possible alternative would be 
to exploit reactions with neutron-rich radioactive beams. Because the intensive 
radioactive beams are not available so far, new isotopes of heaviest nuclei with 
$Z$=111--117 can be synthesized in the $^{48}$Ca-induced  actinide-based complete
fusion-evaporation reactions with the emission of charged particles ($pxn$- and 
$\alpha xn$-channels) from the compound nucleus \cite{hong_17}. The evaporation 
of proton or $\alpha$-particle from compound nucleus in these reactions, for example  
$^{48}$Ca+$^{248}$Cm$\to ^{290}$Fl+$\alpha 2n$, $^{48}$Ca+$^{248}$Cm$\to ^{293}$Mc+$p2n$, 
$^{48}$Ca+$^{249}$Bk$\to ^{294}$Lv+$p2n$, $^{48}$Ca+$^{251}$Cf$\to ^{295}$Ts+$p3n$,  
leads to the formation of nuclei with smaller $Z$, but with  larger neutron excess. 
In addition, in the  nucleus formed the electron capture (EC) can occur by converting 
a proton into a neutron to the daughter nucleus. Therefore, it is of great interest
to study the competition between $\beta^+$/EC and $\alpha$ decays in SHN produced 
in the  $pxn$ and $\alpha xn$ evaporation channels of hot-fusion reactions.

In addition, the $\beta^+$ decay and EC branches would open the possibility to reach 
other SHN not belonging the original $\alpha$-decay chains. These new branches would 
be open if $\beta^+$/EC and $\alpha$-decay half-lives are comparable 
\cite{zagrebaev_12,karpov_12}. In Ref. \cite{hofmann_16}, the competition between 
$\beta^+$/EC and $\alpha$ decays has been considered in $^{290}$Fl, arguing about 
the possibility of populating a new $\alpha$-decay chain started at $^{290}$Nh. 
So, the study of $\beta^+$/EC  decay modes is also important for the unambiguous 
identification of new SHN.

In this paper the focus of attention is the $\beta^+/EC$ decay mode in SHN that so 
far has been studied only at a phenomenological level 
\cite{moller_97,zagrebaev_12,karpov_12,heenen_15}. In this work the $\beta^+/EC$-decay 
half-lives are calculated microscopically from an effective nucleon-nucleon interaction 
within a deformed selfconsistent mean-field Hartree-Fock calculation with Skyrme forces 
and pairing correlations in the BCS approximation. Four SHN are selected as 
representative of this mass region, namely, the isotopes of Flerovium ($Z=114$, $N=176$), 
$^{290}$Fl;  Moscovium ($Z=115$, $N=178$), $^{293}$Mc; Livermorium ($Z=116$, $N=178$), 
$^{294}$Lv; and Tennessine ($Z=117$, $N=178$), $^{295}$Ts.

The structure of the paper is as follows. I first review briefly in Section II the 
theoretical method used to calculate Gamow-Teller (GT) strength distributions and 
$\beta^+/EC$ half-lives. Then, I proceed to show the results in Section III. In 
Section III.A the ability of the method to reproduce the half-lives of nuclei is 
tested in the vicinity of $Z=100$, where experimental data are available. Section 
III.B contains the results for the SHN mentioned above. Finally, Section IV contains 
the summary and conclusions.


\section{Theoretical framework}

The $\beta^+/EC$-decay half-life, $T_{\beta^+/EC}$, is calculated by summing all the 
allowed GT transition strengths to states in the daughter nucleus with excitation 
energies  $E_{ex}$, lying below the corresponding $Q_i$ energy ($i=\beta^+,EC$), 
\begin{equation}
Q_{EC}=Q_{\beta^+} +2m_e= M(A,Z)-M(A,Z-1)+m_e \, , 
\label{qec}
\end{equation}
written in terms of the nuclear masses $M(A,Z)$ and the electron mass ($m_e$). The GT 
strength is weighted with phase-space factors $f^i(Z,W_0)$, where the energy is
$W_0=Q_i-E_{ex}$.
\begin{equation}
T_i^{-1}=\frac{\left( g_{A}/g_{V}\right) _{\rm eff} ^{2}}{D}
\sum_{E_{ex} < Q_{i}}f^i \left( Z,W_0 \right) B(GT,E_{ex}) \, ,
 \label{t12}
\end{equation}
with $D=6143$~s and $(g_A/g_V)_{\rm eff}=0.77(g_A/g_V)_{\rm free}$, where 0.77 is a 
standard quenching factor  and $(g_A/g_V)_{\rm free}=-1.270$.  Forbidden transitions are 
in general much smaller and therefore, they can be safely neglected, especially in 
nuclei with small $Q_i$-energies, such as those studied here.

Therefore, in the calculations of the $\beta^+/EC$ half-lives there are three main 
ingredients: (i) the $Q_{i}$ energies (maximum energy available in the process), 
which are taken from experiment when available or from different mass formulas or 
microscopic calculations in other cases; (ii) the phase-space factors for each 
transition, which are calculated  in a similar way in practically all the existing 
calculations of the half-lives; and (iii) the nuclear structure that generates the energy 
distribution of the GT strength. This distribution may differ much among different 
approaches, such as simple constant values \cite{karpov_12}, distributions calculated 
with phenomenological potentials  \cite{moller_97}, or microscopic calculations based 
on effective nucleon-nucleon interactions like the work presented in this paper.
I specify in what follows how these three pieces are treated.

Some recent calculations of half-lives in SHN  by Karpov {\it et al.} \cite{karpov_12}
assume that the decay can be approximated by considering allowed transitions from the 
ground state of the parent nucleus to the ground state of the daughter. $Q_{i}$ 
energies are taken from the masses of the finite range droplet model (FRDM) \cite{FRDM} 
and the nuclear matrix elements of the transitions are assumed to be constant with
$\log(ft)=4.7$ for all  nuclei. The latter assumption might be very rough because it 
neglects any nuclear structure effect. It could be a large estimate of the average GT 
strength that finally would lead to half-lives being underestimated. In an older paper, 
\cite{fiset_72} the authors used the same approach, but with $\log(ft)=6.5$. Then, in 
those references, only the phase-space factors remain to be calculated. $\beta^+/EC$ 
half-lives were also evaluated within a proton-neutron quasiparticle random phase 
approximation (pnQRPA) approach, which is based on a phenomenological folded-Yukawa 
single-particle Hamiltonian \cite{moller_97}, using masses from FRDM and similar
phase factors. Unfortunately, only $\beta^+/EC$ half-lives smaller than 100 s were 
published and the isotopes studied here are not in this category.

In the calculations of this work, the nuclear structure is described microscopically 
from selfconsistent deformed Hartree-Fock calculations with Skyrme forces and pairing 
correlations. The $Q_{i}$ energies in the cases where the masses are not measured are 
taken from different mass formulas that include masses  from FRDM also used in the 
above references \cite{karpov_12,moller_97}. The calculation of the phase factors is 
similar to those. Therefore, the current calculations represent an improvement over 
the previous ones with respect to the nuclear structure involved in the decay process.


\subsection{Mean-field approach for nuclear structure}

A brief summary of the theoretical formalism used in this paper to describe the 
nuclear structure involved in the $\beta^+/EC$-decay is presented here. Further 
details can be found elsewhere \cite{sarri1,sarri_99,sarri2,sarri_odd}. The starting 
point is a self-consistent calculation of the mean field in terms of a deformed 
Hartree-Fock with Skyrme interactions and pairing correlations in the BCS 
approximation. The Skyrme interaction SLy4 \cite{chabanat} is selected because of 
its ability to account successfully for a large variety of nuclear properties all 
along the nuclear chart \cite{bender_08,stoitsov_03}. Single-particle energies, wave 
functions, and occupation amplitudes are generated in this way. The solution of the 
HF equations is found by using the formalism developed in Ref. \cite{vautherin}, 
assuming time reversal and axial symmetry. The single-particle wave functions are 
expanded into the eigenstates of a harmonic oscillator with axial symmetry in 
cylindrical coordinates, using 16 major shells. 
It is well known that the harmonic oscillator basis used in the expansion of the
deformed Hartree-Fock wave functions exhibits a Gaussian behavior at large distances 
that does not take properly into account effects of the continuum. These effects
may be important in nuclei close to the drip lines. In mean-field
approaches, this problem is cured by using a coordinate representation or a
transformed harmonic oscillator basis that allows one to use a configuration 
space with the correct exponential asymptotic behavior \cite{stoitsov}. Nevertheless, 
continuum effects are not expected to play any important role in the SHN
studied in this work, which are close to islands of stability, and therefore
they can be safely neglected here.

In the mean-field approach, the 
energy of the different shape configurations can be evaluated with constrained 
calculations, minimizing the Hartree-Fock energy under the constraint of keeping 
fixed the nuclear quadrupole deformation. The resulting total energy plots versus 
deformation are called in what follows deformation-energy curves (DEC). Deformation 
has been shown to be a key ingredient to understand the decay properties of 
$\beta$-unstable nuclei  \cite{sarri1,sarri_99,sarri2,sarri_odd} and this would be 
of special importance in SHN. 

In the next step, the GT strengths are calculated for the equilibrium shapes of each 
nucleus, that is, for the  minima obtained in the DECs. Since decays connecting 
different shapes are disfavored, similar shapes are assumed for the ground state of 
the parent nucleus and for all populated states in the daughter nucleus 
\cite{moller1,homma,sarri_pb3}.

To describe GT transitions, a deformed pnQRPA 
\cite{sarri1,sarri_99,sarri2,sarri_odd,moller1,homma,sarri_pb3,hir2}
formalism with spin-isospin  residual interactions is used. However, in SHN the 
coupling strengths of these interactions are expected to be very small because they 
scale with the inverse of the mass number and therefore pnQRPA correlations are not 
expected to be especially relevant here, in particular for the half-lives that are 
only sensitive to the low energy region below the $Q$-window. Then, they are 
neglected in this work. Anyhow, the inclusion of pnQRPA correlations would result 
in a small reduction of the GT strength that would translate into a small increase 
of the corresponding half-lives.

The GT transition amplitudes in the intrinsic frame connecting the ground state 
$| 0^+\rangle $ of an even-even nucleus to one phonon states with energy $\omega_K$ 
in the daughter nucleus $|\omega_K \rangle \, (K=0,1) $ are found to be 
\cite{sarri1,sarri_99,sarri2,sarri_odd,moller1,homma,sarri_pb3,hir2},

\begin{equation}
\left\langle \omega _K | \sigma _K t^{+} | 0 \right\rangle =
\sum_{\pi\nu}\left( \tilde{q}_{\pi\nu}X_{\pi
\nu}^{\omega _{K}}+ q_{\pi\nu}Y_{\pi\nu}^{\omega _{K}} \right) \, ,
\label{intrinsic}
\end{equation}
with
\begin{equation}
\tilde{q}_{\pi\nu}=u_{\nu}v_{\pi}\Sigma _{K}^{\nu\pi },\ \ \
q_{\pi\nu}=v_{\nu}u_{\pi}\Sigma _{K}^{\nu\pi},
\label{qs}
\end{equation}
in terms of the occupation amplitudes for neutrons and protons $v_{\nu,\pi}$   
($u^2_{\nu,\pi}=1-v^2_{\nu,\pi}$) and the matrix elements of the spin operator, 
$\Sigma _{K}^{\nu\pi}=\left\langle \nu\left| \sigma _{K}\right| \pi\right\rangle $, 
connecting proton and neutron single-particle states, as they come out from the 
HF+BCS calculation. $X_{\pi\nu}^{\omega _{K}}$ and $Y_{\pi\nu}^{\omega _{K}}$ are the 
forward and backward amplitudes of the pnQRPA phonon operator, respectively. 

Once the intrinsic amplitudes in Eq. (\ref{intrinsic}) are calculated, the GT 
strength $B(GT^{+})$ in the laboratory system for a transition  
$I_iK_i (0^+0) \rightarrow I_fK_f (1^+K)$ can be evaluated.
Using the Bohr-Mottelson factorization \cite{bm} to express the initial and final 
states in the laboratory system in terms of intrinsic states, one arrives at

\begin{eqnarray}
B(GT^{+} ,\omega )& =& \sum_{\omega_{K}} \left[ \left\langle \omega_{K=0}
\left| \sigma_0t^{+} \right| 0 \right\rangle ^2 \delta (\omega_{K=0}-
\omega ) \right.  \nonumber  \\
&& \left. + 2 \left\langle \omega_{K=1} \left| \sigma_1t^{+} \right|
0 \right\rangle ^2 \delta (\omega_{K=1}-\omega ) \right] \, ,
\label{bgt}
\end{eqnarray}
in $[g_A^2/4\pi]$ units. The strength distributions will be referred to the excitation 
energy in the daughter nucleus, which are given by

\begin{equation}
E_{ex}=\omega -E_{\pi_0}-E_{\nu_0}\, ,
\end{equation}
where $E_{\pi_0}$ and $E_{\nu_0}$ are the lowest quasiparticle energies for protons and 
neutrons, respectively.

To describe odd-$A$ nuclei, I follow the usual strategy of blocking the state 
corresponding to a given $J^{\pi}$ and using the equal filling approximation to 
calculate its nuclear structure \cite{sarri_odd}. This approximation has been compared 
with other more sophisticated approaches, showing that it is sufficiently precise for 
most practical applications \cite{schunck_10}. A microscopic justification has been
given in terms of standard procedures of quantum statistical mechanics \cite{robledo_08}.
In principle, the blocked state is selected  to minimize the energy among the states 
in the vicinity of the Fermi level. In cases where the $J^{\pi}$ of the nucleus is 
experimentally known, the natural option is to choose $J^{\pi}$ according to this value. 
In all the test cases studied later, these states appear always close to the Fermi 
level, as expected. In SHN, where there is no experimental information on $J^{\pi}$, 
the state $J^{\pi}$ that corresponds to the ground state is used, but several choices 
for them among the states that are close to the Fermi level are also used for 
comparison. Studying the sensitivity of the half-lives to the choice of  $J^{\pi}$ is 
interesting because slight changes in the theoretical treatment may lead to different 
$J^{\pi}$ for the ground sates.

The GT strength distributions and $\beta$-decay half-lives have been studied in the 
past within  this model  in various mass regions that include neutron-deficient 
isotopes in the $A\approx 70$ mass region \cite{sarri_wp,sarri_rp} and in the lead 
region \cite{sarri_pb3,sarri_pb1,sarri_pb2}; neutron-rich isotopes in medium-mass  
\cite{sarripere1,sarripere2,sarripere3,kiss}, and rare-earth  nuclei \cite{sarri_rare}; 
and $fp$-shell nuclei \cite{sarri_fp1,sarri_fp2,sarri_fp3}. The sensitivity of the 
GT strength distributions to different ingredients of the theoretical formalism were 
studied in those works with especial emphasis on the deformation dependence of the
decay properties. In particular, the sensitivity of the energy distribution of the 
GT strength to deformation has been exploited to determine the nuclear shape by 
comparing theoretical results with $\beta$-decay measurements using the total 
absorption spectroscopy technique \cite{expnacher}.


\subsection{Phase-space factors}

In $\beta^{+}/EC$ decay, the phase-space factors $f^{\beta^+/EC}( Z,W_0)$ contain two 
parts, positron emission and electron capture. The former, $f^{\beta^+}$, is computed 
numerically for each value of the energy including screening and finite size effects, 
as explained in Ref. \cite{gove},

\begin{equation}
f^{\beta^+} (Z, W_0) = \int^{W_0}_1 p W (W_0 - W)^2 \lambda^+(Z,W) {\rm d}W\, ,
\label{phase}
\end{equation}
with
\begin{equation}
\lambda^+(Z,W) = 2(1+\gamma) (2pR)^{-2(1-\gamma)} e^{-\pi y}
\frac{|\Gamma (\gamma+iy)|^2}{[\Gamma (2\gamma+1)]^2}\, ,
\end{equation}
where $\gamma=\sqrt{1-(\alpha Z)^2}$ , $y=\alpha ZW/p$ , $\alpha$ is the fine 
structure constant, and $R$ is the nuclear radius. $W$ is the total energy of the 
$\beta$ particle, $W_0$ is the total energy available in $m_e c^2$ units, and 
$p=\sqrt{W^2 -1}$ is the momentum in $m_e c$ units.

The electron capture phase factors, $f^{EC}$, have also been included
following Ref. \cite{gove}:

\begin{equation}
f^{EC}=\frac{\pi}{2} \sum_{x} q_x^2 g_x^2B_x \, ,
\end{equation}
where $x$ denotes the atomic sub-shell from which the electron is captured that 
includes $K$- and $L$- orbits. $q$ is the neutrino energy, $g$ is the radial 
component of the bound state electron wave function at the nucleus, and $B$ stands 
for other exchange and overlap corrections \cite{gove}.


\subsection{$Q_{EC}$ energies}

$\beta^+/EC$ half-lives depend critically on the $Q$-energies that determine the 
maximum energy of the transition and the values of the phase factors that weight 
the GT strength, see Eq. (\ref{t12}). In those cases where experimental masses are 
available \cite{audi_12,nndc}, the natural choice is to use these values to evaluate 
Eq. (\ref{qec}). But in those cases where experimental masses are still missing, 
one has to rely on theoretical predictions for them. There are a large number of 
mass formulas in the market obtained from different approaches. The strategy followed 
in this work starts by comparing with experiment the predictions of some representative 
mass formulas in the mass region where data are available and use them later in SHN 
where there is no experimental information.

Among the phenomenological approaches for the masses, I take the FRDM \cite{FRDM} 
that belongs to a macroscopic-microscopic type of calculation. It contains a 
finite-range droplet model corrected by microscopic effects obtained from a deformed
single-particle model based on a folded-Yukawa potential including pairing in the 
Lipkin-Nogami approach. Then, I use the extended Thomas-Fermi plus Strutinsky integral 
(ETFSI) model \cite{ETFSI}, which adopts a semi-classical approximation to the
Hartree-Fock method including full Strutinsky shell corrections and BCS pairing 
correlations. The Duflo and Zuker (DZ) mass model \cite{DufloZuker} is used as well,
which is written as an effective Hamiltonian that contains monopole and multipole 
terms. I also compare with fully microscopic calculations based on effective two-body 
nucleon-nucleon interactions. Among them, I consider the masses from the HFB-21 
model, which is one of the most recent versions of the Skyrme HFB mass formulas 
introduced by the Brussels-Montreal group \cite{HFB21}. I also use the masses 
calculated from the Skyrme forces SkP and  SLy4 with a zero-range pairing force and
Lipkin-Nogami obtained from the code HFBTHO \cite{mass_sly4}. All the masses used 
here can be found in Ref.  \cite{www_masses}.


\section{Results for the half-lives}

\begin{figure}[htb]
\centering
\includegraphics[width=80mm]{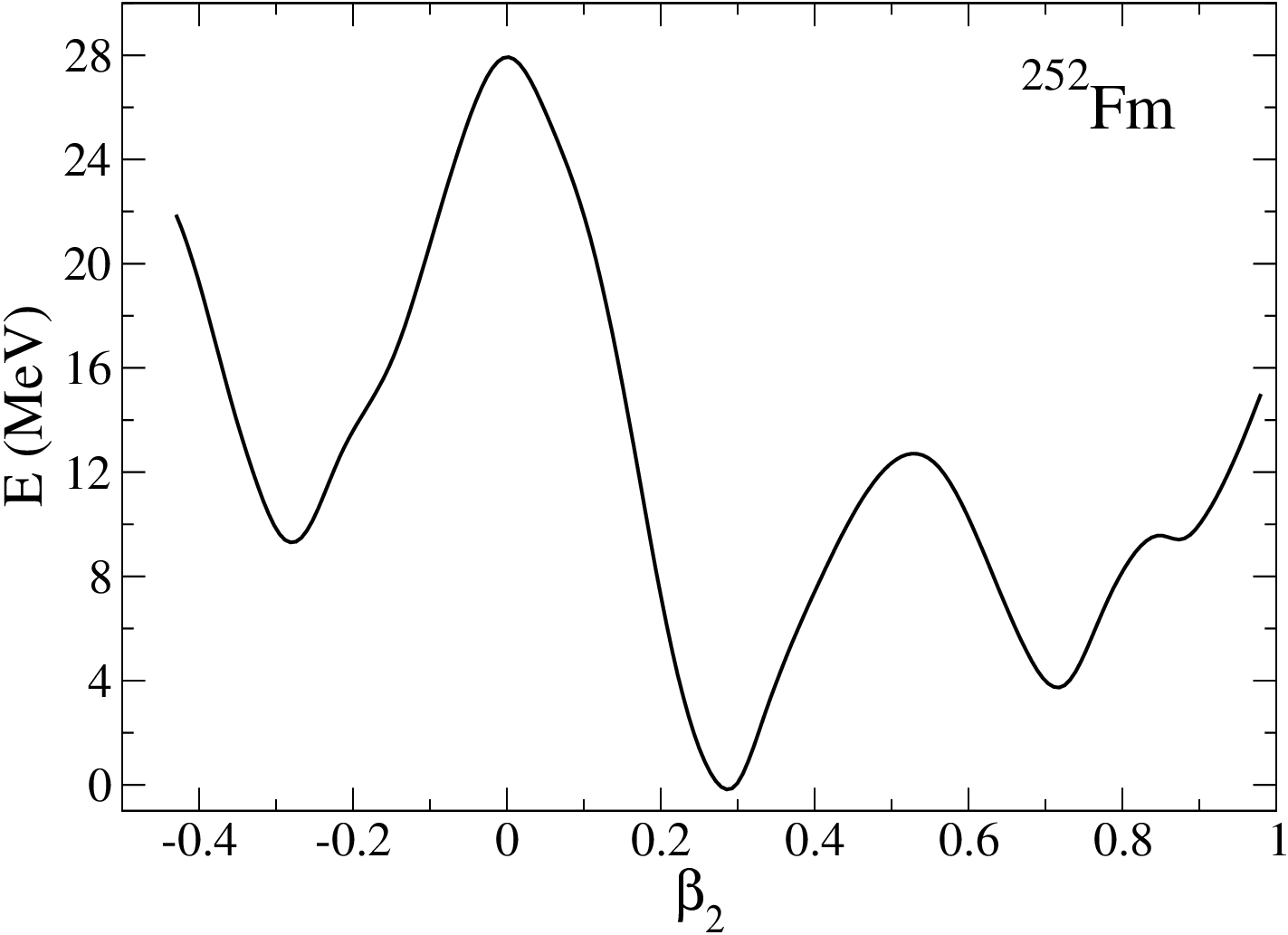}
\caption{Deformation-energy curve for the $^{252}$Fm isotope obtained from 
constrained HF+BCS calculations with the Skyrme force SLy4.}
\label{eq0_I}
\end{figure}

\begin{figure}[htb]
\centering
\includegraphics[width=80mm]{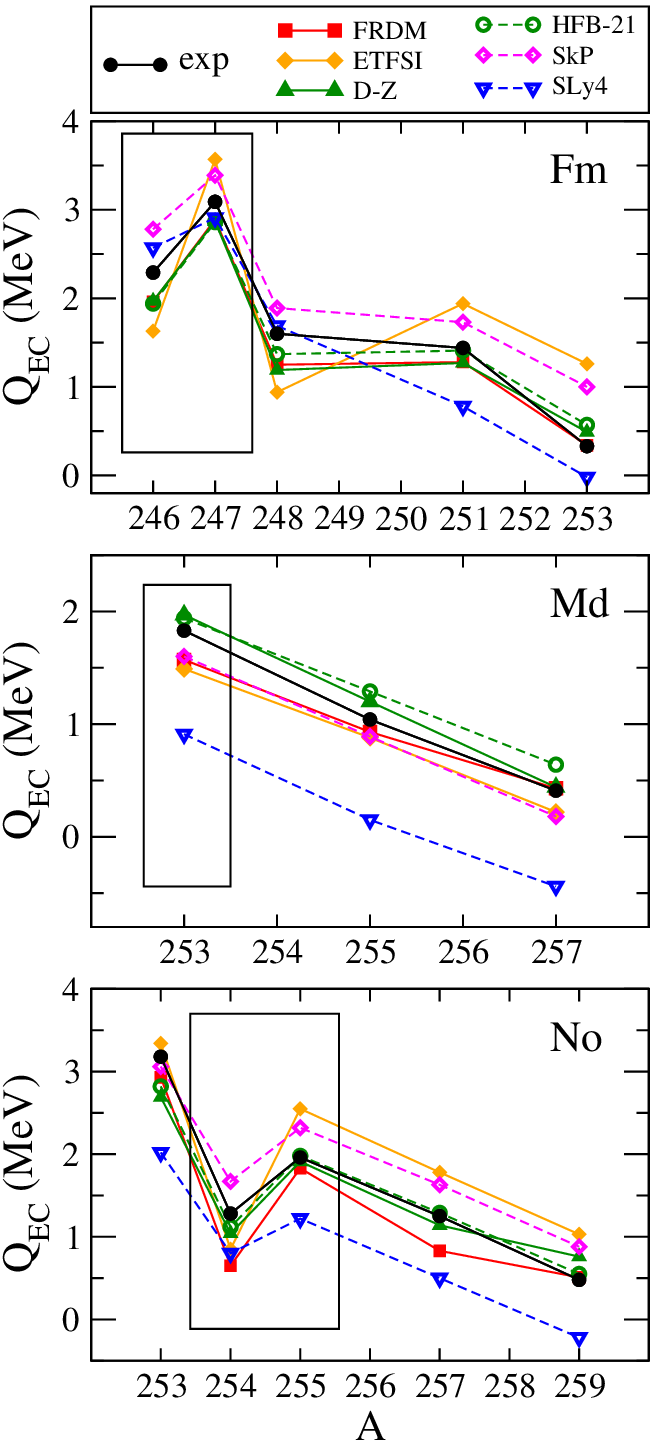}
\caption{$Q_{EC}$ energies (MeV) corresponding to experimental and different 
calculated masses (see text) for Fm, Md, and No isotopes.}
\label{qec_I}
\end{figure}

\begin{figure}[htb]
\centering
\includegraphics[width=80mm]{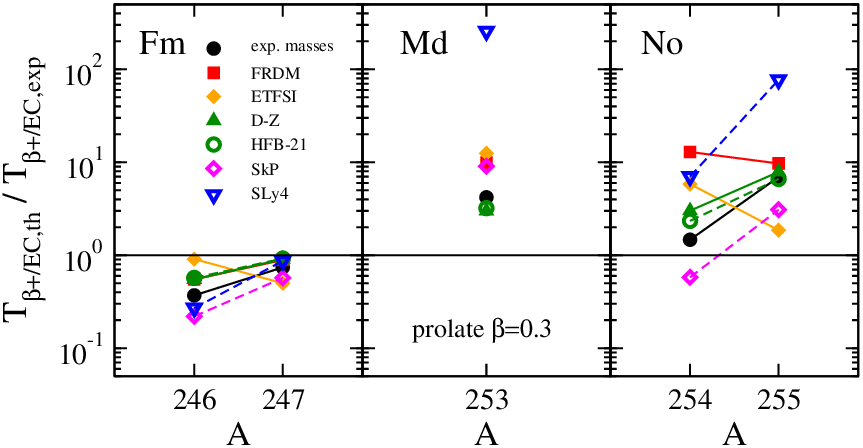}
\caption{Ratios of calculated and experimental  half-lives for Fm, Md, and No isotopes. 
The results correspond to the ground state configurations (prolate $\beta_2 =0.3$) using 
$Q_{EC}$ energies from different mass formulas or microscopic calculations.} 
\label{t12_I_1}
\end{figure}

\begin{figure}[htb]
\centering
\includegraphics[width=80mm]{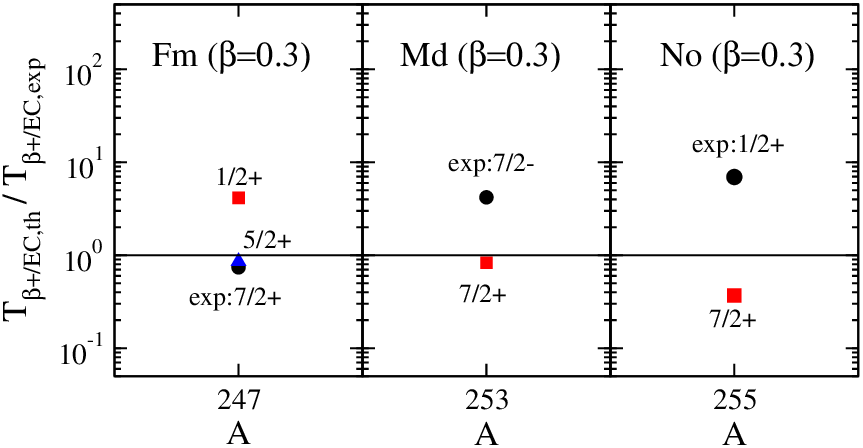}
\caption{Ratios of calculated and experimental half-lives for odd-$A$ Fm, Md, and No 
isotopes. The results correspond to the ground state prolate configurations 
($\beta_2 =0.3$), using the experimental $Q_{EC}$ energies and different $J^{\pi}$ 
assignments for the odd nucleon.} 
\label{t12_I_2}
\end{figure}

\begin{figure}[htb]
\centering
\includegraphics[width=80mm]{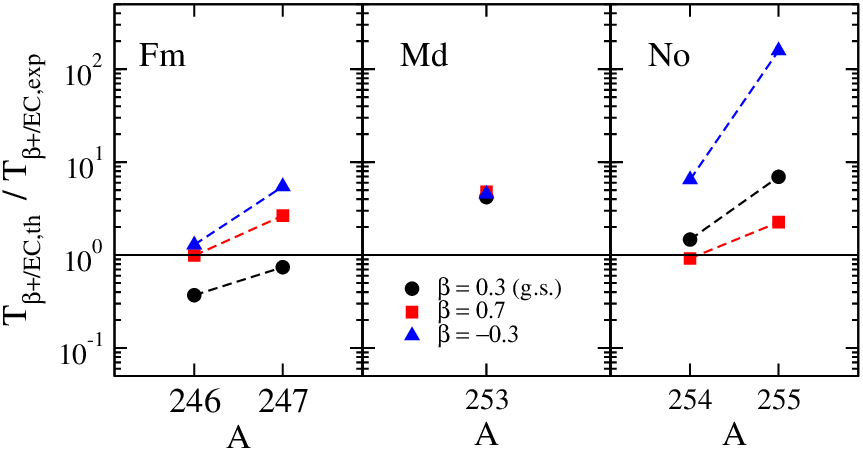}
\caption{Ratios of calculated and experimental half-lives  for Fm, Md, and No isotopes. 
The results are obtained with experimental $Q_{EC}$ energies for the  three shapes 
that produce energy minima in Fig. \ref{eq0_I}.} 
\label{t12_I_3}
\end{figure}

\begin{figure*}[htb]
\centering
\includegraphics[width=150mm]{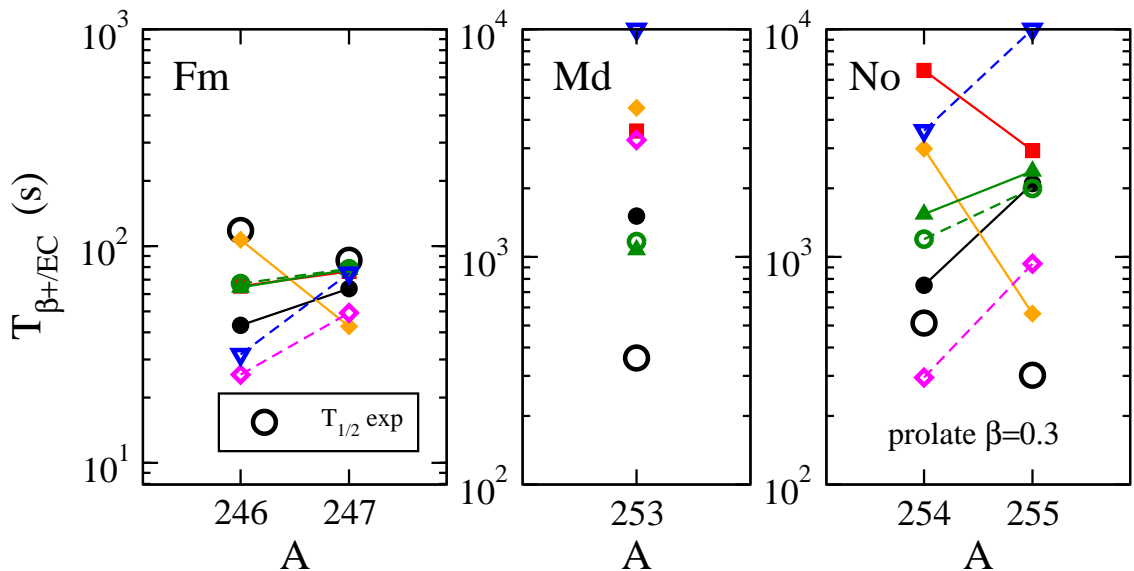}
\caption{Calculated and experimental half-lives for Fm, Md, and No isotopes. The 
results correspond to ground state configurations (prolate $\beta_2 =0.3$) using 
different $Q_{EC}$ energies with the same code symbol of Fig. \ref{t12_I_1}.}
\label{t12_abs_I}
\end{figure*}

In this section I present the calculations for the half-lives in SHN. I first show 
the results obtained for the DECs in the isotopes studied. The energy distributions 
of the GT strength corresponding to the local minima of the DECs are calculated 
afterwards. Finally, half-lives are computed.

Before starting with the calculations of the SHN mentioned in the Introduction, the 
quality of the calculations is checked in some isotopes around $Z=100$, where both  
$Q_{EC}$ and $T_{\beta^+/EC}$ have been measured. Namely, the isotopes of Fermium 
($Z=100$) $^{246,247}$Fm;  Mendelevium ($Z=101$) $^{253}$Md; and Nobelium ($Z=102$) 
$^{254,255}$No, are considered. After this test, the results for the SHN studied in 
this work, $^{290}$Fl, $^{293}$Mc, $^{294}$Lv, and $^{295}$Ts, will be shown.


\subsection{Testing case: Fm, Md, and No isotopes}

Studying the deformation dependence of the energy by constrained calculations shows 
that  nuclei in this region present three minima that correspond to oblate, prolate, 
and large prolate shapes. In Fig. \ref{eq0_I}  the DEC for $^{252}$Fm with SLy4 is 
shown. The energies are relative to the ground state energy, as a function of the 
quadrupole deformation $\beta_2$. These results are very similar to the DECs for the
other testing isotopes and I discuss only this case as an example. The ground state 
is found to have a prolate shape with a quadrupole deformation around 
$\beta_2 \approx 0.3$, but 
there are also minima at oblate $\beta_2 \approx -0.3$ and prolate $\beta_2 \approx 0.7$ 
configurations at typical excitation energies around 8 and 4 MeV, respectively. 
These results agree quite well with calculations performed with the finite-range 
Gogny D1S interaction \cite{gogny}.

Experimental masses of parent and daughter nuclei in this mass region are available 
and then, one gets experimental $Q_{EC}$ energies. However, in the heavier nuclei 
considered in the next section, this information is missing and one has to rely on the 
predictions of mass  formulas. I have  considered some of the most commonly used 
formulas (or microscopically calculated masses) as they appear in the nuclearmasses.org 
web page \cite{www_masses}. They are FRDM, ETFSI, DZ, HFB-21, SkP, and SLy4, introduced 
in the previous section. Figure \ref{qec_I} shows  the $Q_{EC}$ energies from experiment 
and from different mass formulas and  illustrates the spreading of the $Q_{EC}$ energies.
The cases for which half-lives have been calculated appear within a frame in 
Fig. \ref{qec_I}. One can see that the results are scattered about 1 MeV between the 
largest and smallest energies among the cases considered. Experimental values appear 
within these extreme values. This gives us a fair idea of the uncertainties expected. 
Although the uncertainty is not very large, these $Q$-values determine the energy 
range of excitations that contribute to the half-lives, as well as the magnitude of 
the phase factors and as it will be seen in the next figures, the effect on the 
half-lives is important.

Figures \ref{t12_I_1}-\ref{t12_I_3}  show the ratios of the calculated half-lives 
to the experimental ones for some Fm, Md, and No isotopes, where there are experimental 
data. The experimental values have been extracted from the total half-lives measured 
together with the percentage that corresponds to the $\beta^+/EC$ decay. 
Figure \ref{t12_I_1} shows  the results for the ground states (prolate with 
$\beta_2 \approx 0.3$) and using the experimental $J^{\pi}$ in the case of odd-$A$ nuclei.
The various calculations correspond to the different  $Q_{EC}$ values either from
experiment or from calculated masses. The half-lives of Fm isotopes are underestimated, 
whereas the half-lives of Md and No isotopes are somewhat overestimated. One can see 
a clear correlation between the $Q_{EC}$ energies in Fig. \ref{qec_I} and the half-lives 
in Fig. \ref{t12_I_1}, that is, half-lives decrease with increasing values of $Q_{EC}$.

The decay would be in principle from the ground state of the parent nucleus (that 
determines the shape and $J^{\pi}$ in odd-$A$ nuclei), but I also performed calculations 
of $\beta^+/EC$ half-lives that correspond not only to the ground states, but also to 
other shapes and $J^{\pi}$. This helps us to understand the sensitivity of the results 
to different factors arising from various uncertainties. 

Figure \ref{t12_I_2} shows the sensitivity of the half-lives to the $J^{\pi}$ assignments 
in odd-$A$ nuclei. The results are for the ground state shapes and experimental $Q_{EC}$ 
values. The odd states are chosen according to the experimental spin and parity, as 
well as other possibilities for states that appear very close to the Fermi level.
In Fig. \ref{t12_I_3} one can see the results obtained with experimental $Q_{EC}$
values, but for different shapes that include the ground state ($\beta_2 \approx 0.3$), 
as well as the oblate ($\beta_2 \approx -0.3$) and superdeformed prolate
($\beta_2 \approx 0.7$), depicted in Fig. \ref{eq0_I}. Finally, Fig. \ref{t12_abs_I} 
compares the  half-lives (seconds) measured with the calculated ones using the ground 
state deformations and different prescriptions for the $Q_{EC}$ energies. This figure 
is similar to Fig. \ref{t12_I_1} but for the absolute values.

From the results for the half-lives,  one can learn about  the uncertainties associated 
with different aspects of the calculations. The uncertainties on the half-lives related 
to the $Q_{EC}$ energies, $J^{\pi}$ assignments, and nuclear shapes are comparable, 
spreading the results about one order of magnitude. The agreement with experiment is 
roughly within this order.

\begin{figure}[htb]
\centering
\includegraphics[width=80mm]{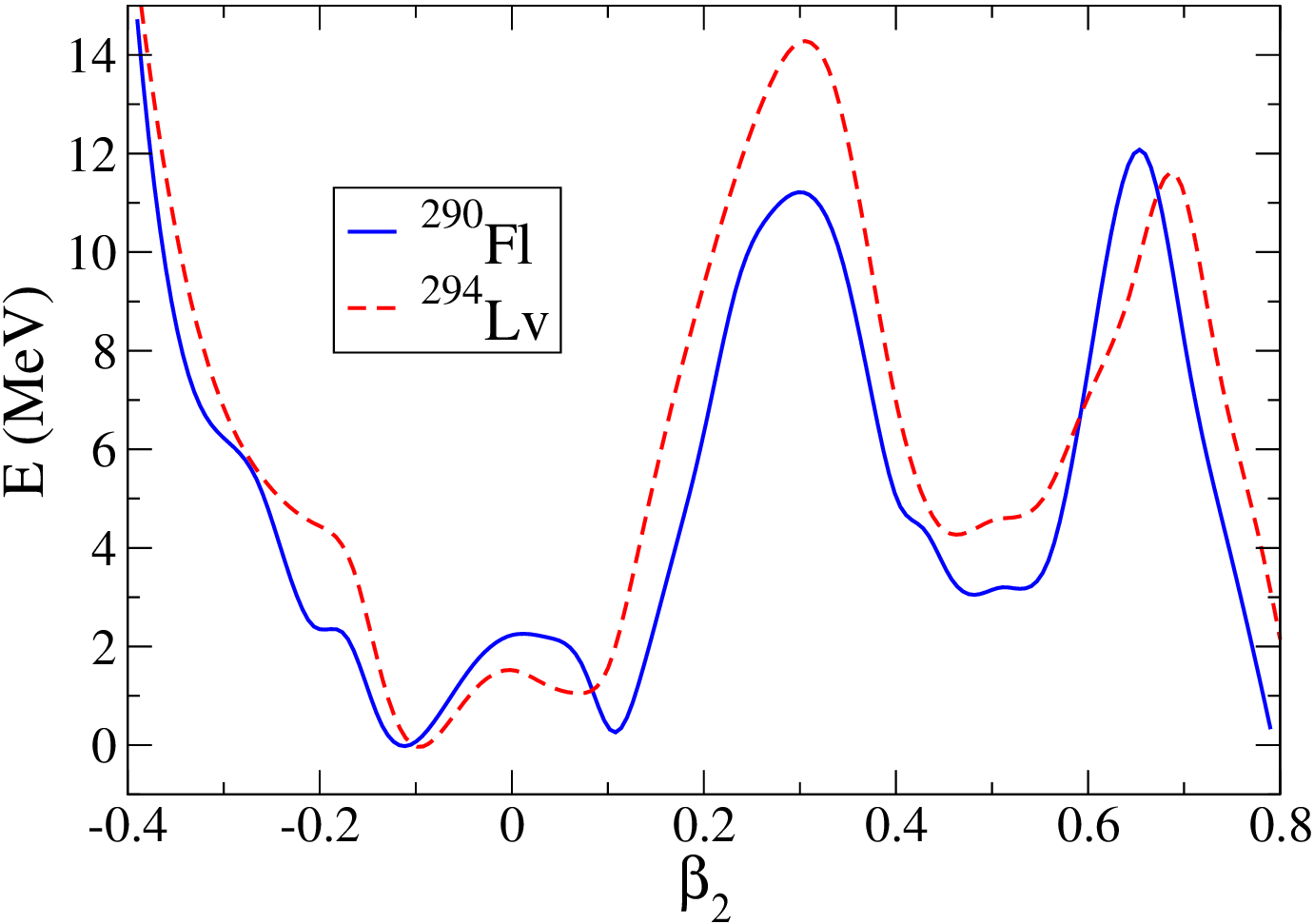}
\caption{Same as in Fig. \ref{eq0_I}, but for$^{290}$Fl and $^{294}$Lv.}
\label{eq0_II}
\end{figure}

\begin{figure}[htb]
\centering
\includegraphics[width=80mm]{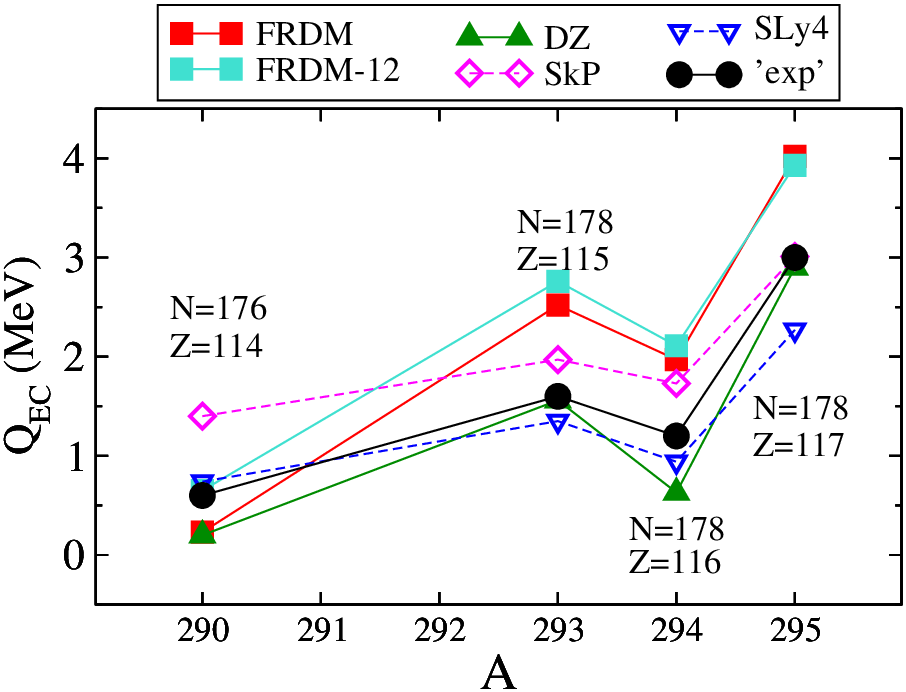}
\caption{$Q_{EC}$ energies (MeV) corresponding to different calculated masses
for SHN.}
\label{qec_II}
\end{figure}

\begin{figure}[htb]
\centering
\includegraphics[width=80mm]{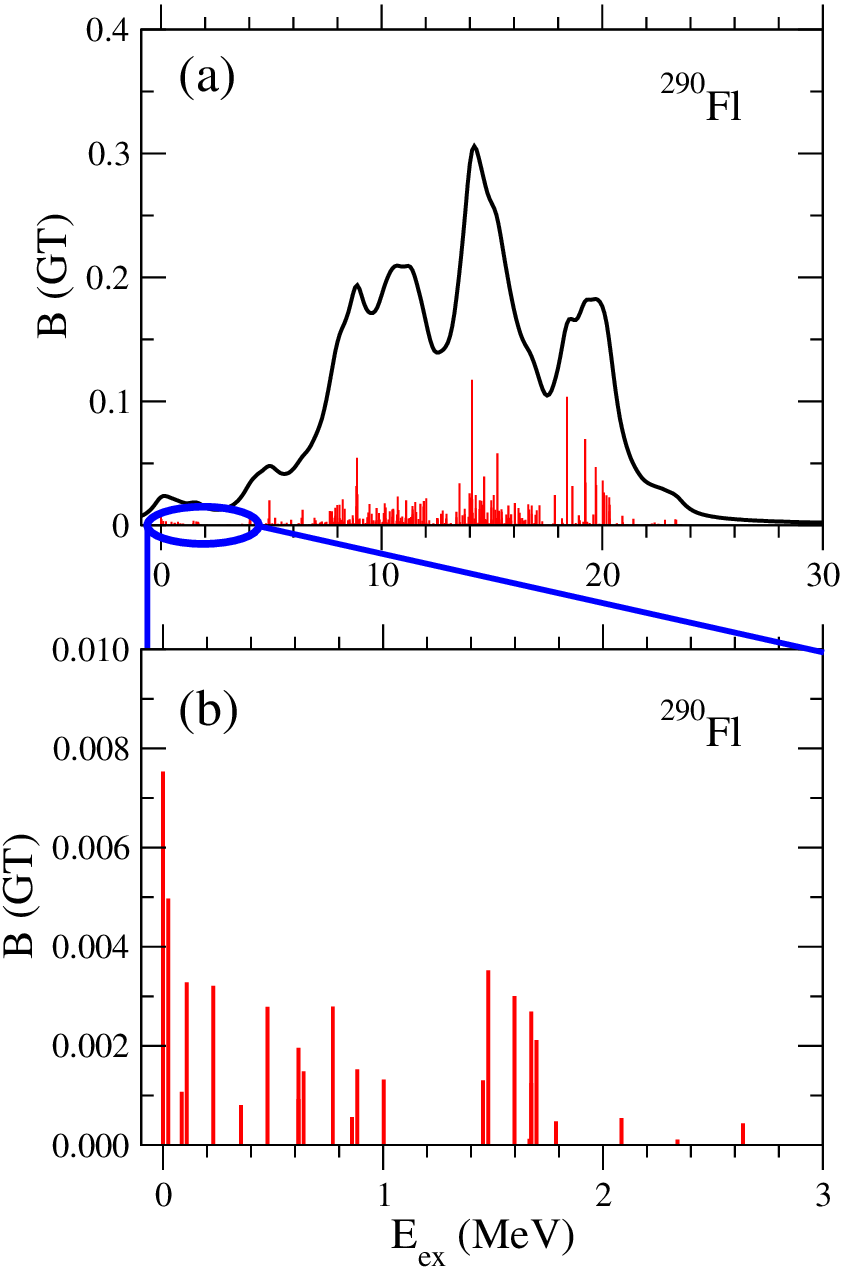}
\caption{(a) Gamow-Teller strength distribution in $^{290}$Fl in the whole range of 
excitation energies of the daughter nucleus.
(b) Magnified region below 3 MeV that includes the $Q_{EC}$ energy.}
\label{bgt_114_176}
\end{figure}

\begin{figure}[htb]
\centering
\includegraphics[width=80mm]{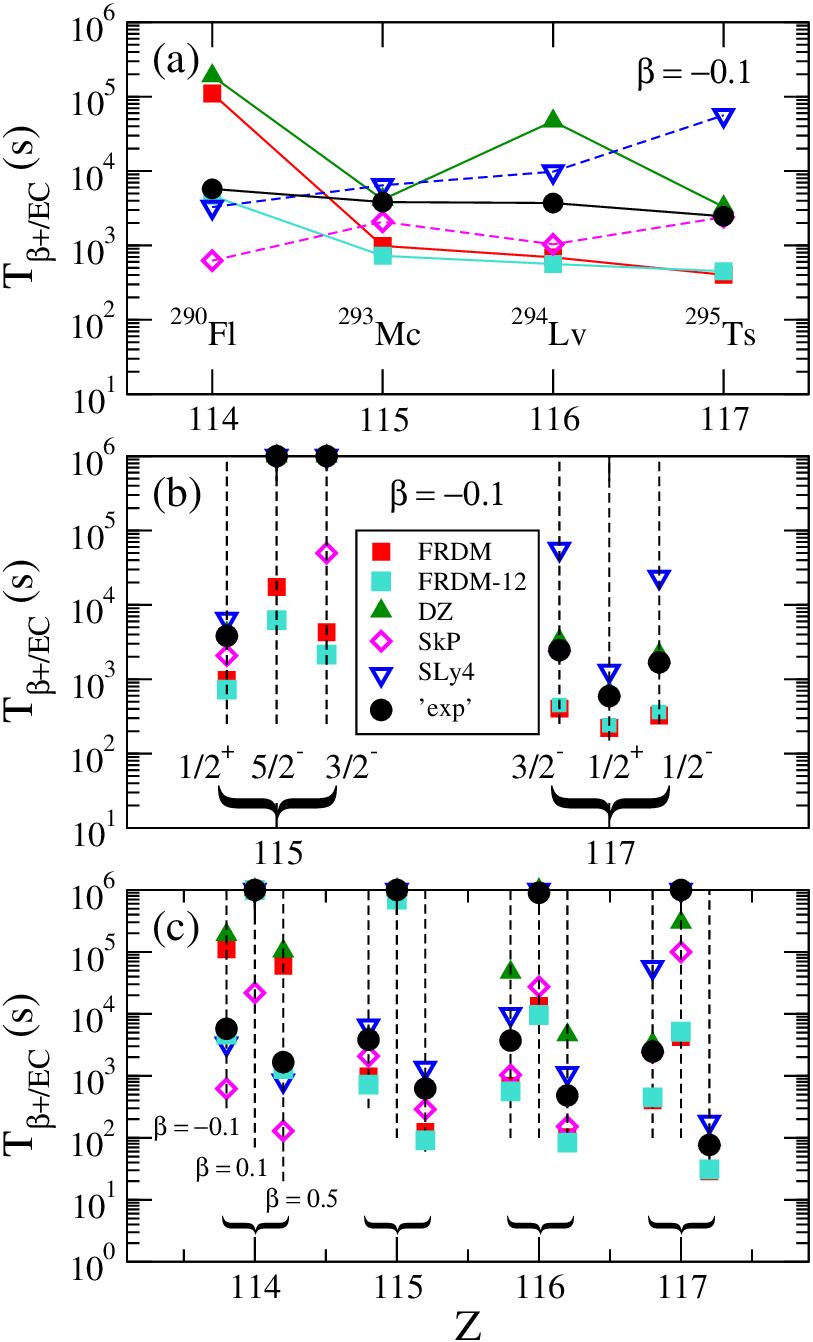}
\caption{Calculated half-lives, $T_{\beta^+/EC}$ (s), for $^{290}$Fl, $^{293}$Mc, $^{294}$Lv,
and $^{295}$Ts. (a) Half-lives for ground state configurations (oblate $\beta_2 =-0.1$) 
with various mass formulas to calculate $Q_{EC}$ energies. (b) Half-lives of odd-$A$ 
nuclei for ground state configurations (oblate $\beta_2 =-0.1$) and several choices of 
$J^{\pi}$ values with various mass formulas to calculate $Q_{EC}$ energies. (c) Half-lives 
for oblate $\beta_2 =-0.1$ (left vertical lines), prolate $\beta_2 =0.1$ (middle vertical 
lines), and prolate $\beta_2 =0.5$ (right vertical lines) nuclear shapes obtained from 
various mass formulas to calculate $Q_{EC}$.}
\label{t12_abs_II}
\end{figure}


\subsection{Superheavy nuclei}

In this section the results for  $^{290}$Fl, $^{293}$Mc, $^{294}$Lv, and $^{295}$Ts are 
discussed. In Fig. \ref{eq0_II}  the plots of the  DECs are shown, relative to the 
ground state energy, for $^{290}$Fl and  $^{294}$Lv as a function of the quadrupole 
deformation $\beta_2$ obtained from the Skyrme force SLy4. In both cases, the ground 
state is the oblate solution ($\beta_2 \approx -0.1$), while two more prolate minima 
appear at $\beta_2 \approx 0.1$ and $\beta_2 \approx 0.5$ at excitation energies of 
about 1 MeV and 4 MeV, respectively. These results agree with those obtained from 
the Gogny-D1S interaction \cite{gogny}.

Figure \ref{qec_II} shows the $Q_{EC}$ energies calculated  with masses obtained from 
three mass formulas (FRDM, FRDM-12, and DZ) and two microscopic calculations (SkP 
and SLy4). The masses from FRDM-12 \cite{moller_12} are a recent improved upgrade of 
the FRDM masses. Although there are no experimental values for these nuclei, I also 
add extrapolated values from the systematics in this mass region extracted from Ref.
\cite{nndc}, that appear in the figure  under the label 'exp'. Similarly to the 
$Q_{EC}$ energies in Fig. \ref{qec_I}, the results in Fig. \ref{qec_II} are distributed 
within 1 MeV with the 'exp' values lying inside this range. 

As a general comment, it is worth noting that the typical $Q_{EC}$ energies in these
nuclei are rather small and, as a consequence, the half-lives are only sensitive
to a very tiny part of the whole GT response of the nucleus. This means also that 
small changes in the nuclear structure description or in the $Q_{EC}$ energies, may 
produce very large effects on the half-lives. This is illustrated in 
Fig. \ref{bgt_114_176} for $^{290}$Fl, where one can see the energy distribution of the 
GT strength in the whole range of energy (a) and below the 3 MeV window (b), where the 
different mass models predict the $Q_{EC}$ energy. The half-life is only sensitive to 
the strength distribution in this small window.

The results for the  $\beta^+/EC$-decay half-lives of the SHN are shown in 
Fig. \ref{t12_abs_II}. The top  figure (a) summarizes the results. They correspond to 
the half-lives for the ground state configuration (oblate $\beta_2\approx -0.1$). The 
states $J^{\pi}$ in the odd-$A$ nuclei are those that minimize the energy. The 
spreading of the results corresponds to the different  $Q_{EC}$ prescriptions and 
there is a clear correlation between the $Q_{EC}$ energies in Fig. \ref{qec_II} and 
the half-lives obtained with them. Thus, the large values of $Q_{EC}$ with the masses 
from FRDM and SkP make the half-lives shorter, whereas DZ and SLy4 having smaller 
$Q_{EC}$ energies, lead to larger  half-lives. The half-lives obtained with the $Q_{EC}$ 
energies extrapolated from the experimental energies in neighbor nuclei appear around 
the average values.

In the middle figure (b) one can see the different results for odd-$A$ nuclei, using 
other $J^{\pi}$ states, which are also close to the Fermi level. The ground state of 
$^{293}$Mc ($Z=115$) corresponds to a $1/2^+$ state that originates from the 
$i_{\rm 13/2}$ spherical orbital and there are two states very close in energy that 
correspond to $5/2^-$ from  $f_{\rm 5/2}$ and  $3/2^-$ from  $p_{\rm 3/2}$. Similarly, 
the ground state of $^{295}$Ts ($Z=117$) corresponds to a $3/2^-$ state whose origin 
is at $p_{\rm 3/2}$ spherical orbital, while two states very close in energy appear 
at $1/2^+$ ($i_{\rm 13/2}$) and  $1/2^-$ ($p_{\rm 3/2}$). The sensitivity of the results 
to the $J^{\pi}$ assumed in the parent nucleus can be understood from the 
characteristics of the allowed transitions considered in this work. Allowed transitions 
correspond to $\Delta \pi =0$ and $\Delta J =0,\pm 1$ transitions and because of the 
small $Q_{EC}$ energies involved, only the low-lying excitations connecting the odd 
proton in $Z=115, 117$ nuclei with neutron states in the vicinity of the Fermi level 
obeying the above selection rules are relevant. In the case of $Z=115$, it turns out 
that in the energy region around the neutron Fermi level, most of the states are 
positive-parity states and then, transitions from a proton $1/2^+$ states are favored, 
whereas transitions from $3/2^-$ and $1/2^-$ states  are suppressed. In the case of 
$Z=117$ the situation is similar, but some $1/2^-$ neutron states are now present 
close to the neutron Fermi level. Then, although transitions from $1/2^+$ states are 
still stronger, decays from $1/2^-$ and $3/2^-$ are not so different. This explains 
why the half-lives of the positive-parity states $(1/2^+)$ are smaller than those of 
the negative-parity states.

In the bottom plot (c) one can see the results corresponding to the oblate 
($\beta_2 \approx -0.1$) in the left vertical lines, prolate ($\beta_2\approx 0.1$) in 
the middle vertical lines, and superdeformed prolate ($\beta_2\approx 0.5$) 
configurations 
in the right vertical lines for each nucleus. The deformations correspond to the minima 
of the DECs in Fig. \ref{eq0_II}. The half-lives obtained from the prolate shapes with 
$\beta_2\approx 0.1$ are in general larger than the ground state oblate values and then, 
they will not play any role in the decay. On the other hand, according to our 
calculations, the half-lives of the superdeformed shape isomers ($\beta_2\approx 0.5$) 
are reduced by about one order of magnitude with respect to those of the ground states.  
The ground states of the superdeformed odd isotopes turn to be $9/2^+$ states 
($i_{\rm 13/2}$), which are very different from the  $J^{\pi}$ of the oblate ground states, 
thus favoring the shape isomeric  possibility of these states. Using the extrapolated 
experimental $Q_{EC}$ energies, half-lives in the range of 100-1000 s in $^{290}$Fl, 
$^{293}$Mc, and $^{294}$Lv and around 100 s in $^{295}$Ts are obtained for the superdeformed 
shapes. The possible decays from these shape isomers might compete with $\alpha$-decays, 
as will be seen in the next section. Because of the rather small excitation energies 
of these states of about 4 MeV according to Fig. \ref{eq0_II}, they could be populated 
in hot-fusion reactions.

Summarizing these results, one can say that the  $\beta^+/EC$-decay half-lives
obtained for the SHN $^{290}$Fl, $^{293}$Mc, $^{294}$Lv, and $^{295}$Ts, vary from  
500 s up to $10^5$ s, depending on the $Q_{EC}$ energies used. The average values 
around  $10^3-10^4$ s, are compatible with the the values obtained from the 
experimental extrapolated values for $Q_{EC}$. Decays from superdeformed shapes 
reduce the half-lives, making them comparable to $\alpha$-decays in some cases.

\begin{figure*}[htb]
\centering
\includegraphics[width=150mm]{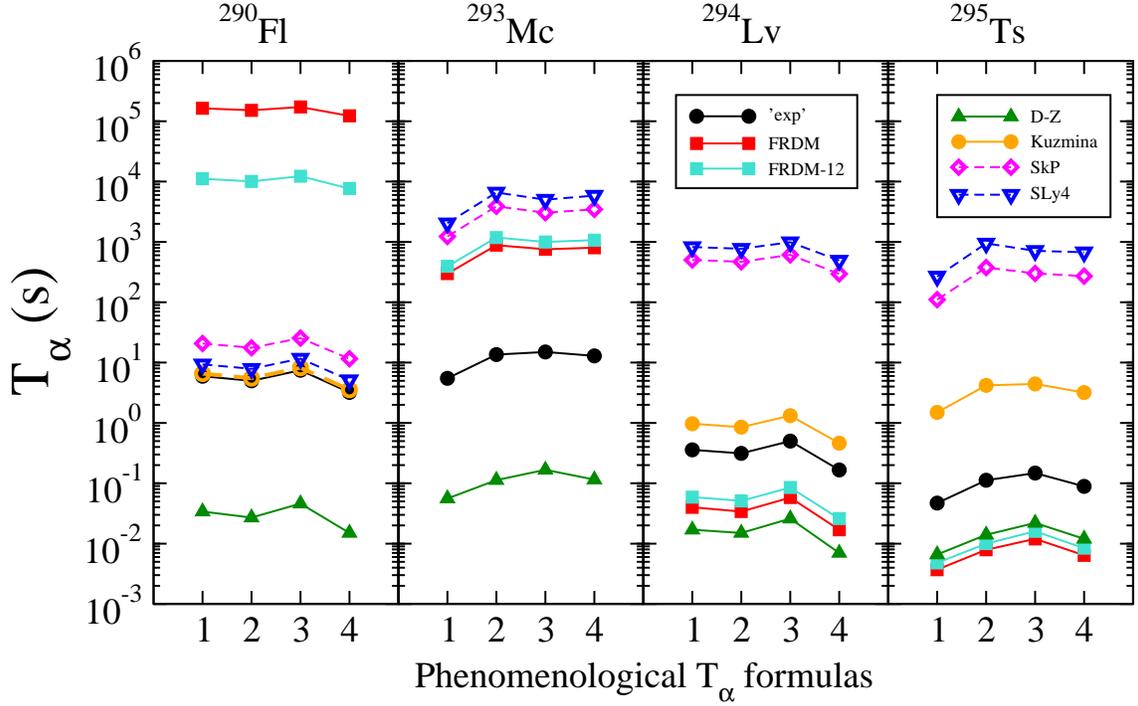}
\caption{$\alpha$-decay half-lives, $T_{\alpha}$ (s), for $^{290}$Fl, $^{293}$Mc, $^{294}$Lv, 
and $^{295}$Ts. The calculations correspond to six different choices for $Q_{\alpha}$ 
values, which are labeled in the figure, and four different options for phenomenological 
formulas of $T_{\alpha}$, labeled from 1 up to 4 in the x-axis (see text).}
\label{talfa}
\end{figure*}


\subsection{$\alpha$-decay half-lives}

As already mentioned in the Introduction, the competition between the different decay 
modes is important to determine the pathways through stability and the SHN that can 
be reached from a given hot-fusion reaction. Therefore, I calculate in this section 
$\alpha$-decay half-lives to be compared with the corresponding $\beta^+/EC$-decay 
half-lives of the previous section.

Similarly to the case of the $Q_{EC}$ energies discussed above, the $Q_{\alpha}$ energies 
can be obtained from the same mass evaluations used for $Q_{EC}$, using the expression

\begin{equation}
Q_{\alpha}  = M(A,Z)-M(A-4,Z-2)-M(4,2)  \, ,
\end{equation}
written in terms of the nuclear masses $M(A,Z)$.  In addition, I also include the 
values calculated in  Ref.  \cite{kuzmina_12}, obtained from a macroscopic-microscopic 
approach based on the two-center shell model applied to superheavy elements. There are 
no experimental measured values for these nuclei yet, but I quote in the last column 
the values obtained from extrapolation of the measured $Q_{\alpha}$ energies in 
neighboring nuclei \cite{nndc}. These values are shown in Table \ref {table_qalfa}. 
They are representative of the most commonly used mass evaluations, but even more 
calculations of $Q_{\alpha}$ energies can be found in the review of Ref. \cite{heenen_15}, 
where values from models based on Woods-Saxon, SkM*, generator coordinate method, and 
relativistic Hartree-Bogoliubov are given as well. The values in Table \ref {table_qalfa}
are in general agreement with those in  Ref. \cite{heenen_15}.


\begin{table}[h]
\caption{$Q_\alpha$ energies (MeV) obtained from different mass models. The energies 
in the last column 'exp' correspond to an extrapolation of the experimental values 
\cite{nndc}.} 
{\begin{tabular}{cccccccc} \hline \hline
Nucleus  & FRDM & FRDM-12 &DZ & SkP  & SLy4 & \cite{kuzmina_12} & 'exp' \\
\hline
$^{290}$Fl   &   8.50 &  8.84 & 10.76 & 9.72 &  9.84 & 8.90     & 9.9  \\
$^{293}$Mc   &   9.47 &  9.44 & 10.84 & 9.28 &  9.21 &  -        & 10.1 \\
$^{294}$Lv   & 10.97 &  10.91 & 11.13 & 9.48 &  9.41 & 10.44  & 10.6 \\
$^{295}$Ts   & 11.58 & 11.54 &  11.48 & 9.85 &  9.72 & 10.53  & 11.1  \\
\hline \hline
\label{table_qalfa}
\end{tabular}}
\end{table}

The $\alpha$-decay half-lives of these nuclei are not measured yet, but there exist in 
the literature phenomenological formulas that have been fitted in different mass 
regions and that can be very useful to see the systematics and to predict these values
in other regions not yet measured. I present here four of these parametrizations, 
which are specifically designed to account for the properties of SHN. These are 
the following: 

\begin{itemize}
\item The formula by  Parkhomenko and Sobiczewski \cite{parkhomenko_05}
(label 1 in the x-axis of Fig.\ref{talfa}):
\begin{equation}
\log_{10} (T_{\alpha})=a Z  (Q_{\alpha}-E_{\mu})^{-1/2} + b  Z + c \, ,
\end{equation}
with  $a=1.5372$, $b=-0.1607$, $c=-36.573$, $E_\mu \rm{(even-even)}$=0, 
$E_\mu$(odd-proton) = 0.113 MeV.

\item The formula by Royer \cite{royer_00}
(label 2 in the x-axis of Fig.\ref{talfa}):
\begin{equation}
\log_{10} (T_{\alpha})=a Z  (Q_{\alpha})^{-1/2} + b  Z^{1/2} A^{1/6}  + c \, ,
\end{equation}
with parameters from Ref. \cite{parkhomenko_05} for even-even nuclei, $a=1.5519$, 
$b=-0.9761$, $c=-28.688$, and for odd-proton nuclei, $a=1.6070$, $b=-0.9467$, $c=-30.912$.

\item The Viola-Seaborg formula \cite{viola_66}
(labels 3 and 4 in the x-axis of Fig.\ref{talfa}):
\begin{equation}
\log_{10} (T_{\alpha})=(a Z + b) (Q_{\alpha})^{-1/2} + (c  Z + d) + h_i \, .
\end{equation}
Two different sets of parameters are used for this formula:

(label 3)  \cite{parkhomenko_05},  
$a=1.3892$, $b=13.862$, $c=-0.1086$, $d=-41.458$, $h_{ee}=0$, $h_{\rm{odd-proton}}=0.437$,
and 

(label 4) \cite{karpov_12,sobiczewski_89}, 
$a=1.66175$, $b=-8.5166$, $c=-0.20228$, $d=-33.9069$, $h_{ee}=0$, $h_{\rm{odd-proton}}=0.772$.
\end{itemize}

Figure \ref{talfa} shows the  $\alpha$-decay half-lives $T_{\alpha}$ (s) for  $^{290}$Fl, 
$^{293}$Mc, $^{294}$Lv, and $^{295}$Ts. The results correspond to the seven different 
choices for $Q_{\alpha}$ given in Table \ref{table_qalfa} and four different options 
for phenomenological formulas of $T_{\alpha}$, labeled from 1 up to 4 in the x-axis as 
explained above. These results agree with similar calculations performed in 
Refs. \cite{heenen_15,moller_97}.

From this figure one can see  that, whereas phenomenological formulas for $T_{\alpha}$
give quite similar results, a strong dependence on the $Q_{\alpha}$ energies is found. 
$T_{\alpha}$ can vary as much as five orders of magnitude (even more in $^{290}$Fl) due to 
the uncertainties in $Q_{\alpha}$.

Phenomenological mass formulas, such as DZ and FRDM, have a tendency to predict short 
values of $T_{\alpha}$, which is a consequence of the large $Q_{\alpha}$ values (see 
Table \ref {table_qalfa}). The exception is FRDM in $^{290}$Fl that predicts the 
largest $T_{\alpha}$ value. On the other hand, microscopic mean-field calculations with 
Skyrme forces (SkP and SLy4) predict larger $T_{\alpha}$ values in these nuclei. The 
$T_{\alpha}$ obtained from the macroscopic-microscopic approach of Ref. \cite{kuzmina_12} 
are close to the half-lives calculated from the inferred experimental values. They 
represent a kind of average value that can be taken as a reference value to compare 
with the $T_{\beta}$ decays. Thus, $T_{\alpha}$ half-lives of the order of 10 s are 
expected in $^{290}$Fl, from 1 to 10 s in $^{293}$Mc, from 0.1 to 1 s in $^{294}$Lv, 
and from 0.01 to 1 s in $^{295}$Ts. These values are always lower than the corresponding 
$T_{\beta^+/EC}$ half-lives, and therefore $\beta^+/EC$ decay would be much slower than 
$\alpha$ decay in these nuclei, not competing with them. Only the $\beta^+/EC$ decay 
from superdeformed shapes with $T_{\beta^+/EC}$ half-lives around 10-100 s could have a 
chance to compete with $\alpha$ decay.


\section{Summary and conclusions}

In this paper $\beta^+/EC$-decay half-lives in $^{290}$Fl, $^{293}$Mc, $^{294}$Lv, and 
$^{295}$Ts, which are representative of SHN created in hot-fusion reactions, have been 
calculated microscopically. The calculations are based on a deformed Skyrme HF+BCS 
approach.

The uncertainties in the $\beta^+/EC$-decay half-lives that originate from poorly known 
$Q$-energies and $J^{\pi}$ assignments, as well as the influence of deformation, have 
been studied. The results are compared with $\alpha$-decay half-lives obtained from 
phenomenological parametrizations using the same mass formulas to determine the 
$Q_{\alpha}$ values. 

Taking into account all the uncertainties in the results from both  $\alpha$ and 
$\beta^+/EC$ decays, it is found that the latter are much larger than the former and 
therefore, there is in general no room for $\beta^+/EC$ decay to play a role in the 
decays of SHN produced in these hot-fusion reactions. The only possibility for a 
competition between both modes of decay would be the decay from superdeformed shape 
isomers that might be populated in the reactions.


\begin{acknowledgments}
I would like to thank G. Adamian for calling my attention to this problem, as well 
as for valuable discussions and a careful reading of the manuscript. This work was 
supported by Ministerio de Ciencia, Innovaci\'on y Universidades under Contract No. 
PGC2018-093636-B-I00.  
\end{acknowledgments}




\begin{thebibliography}{99}
\bibitem{hofmann_00} S. Hofmann and G. M\"unzenberg, 
Rev. Mod. Phys. {\bf 72}, 733 (2000).

\bibitem{oganessian_04} Yu. Ts. Oganessian, V. K. Utyonkov, Yu. V. Lobanov, 
F. Sh. Abdullin, A. N. Polyakov, I. V. Shirokovsky, Yu. S. Tsyganov,
G. G. Gulbekian, S. L. Bogomolov, A. N. Mezentsev, S. Iliev, V. G. Subbotin, 
A. M. Sukhov, A. A. Voinov, G. V. Buklanov, K. Subotic, V. I. Zagrebaev, 
M. G. Itkis, J. B. Patin, K. J. Moody, J. F. Wild, M. A. Stoyer, N. J. Stoyer, 
D. A. Shaughnessy, J. M. Kenneally, and R. W. Lougheed, 
Phys. Rev. C {\bf 69}, 021601(R) (2004).

\bibitem{oganessian_07} Yuri Oganessian,  
J. Phys. G: Nucl. Part. Phys. {\bf 34}, R165 (2007).

\bibitem{oganessian_15} Yu. Ts. Oganessian and V. K. Utyonkov, 
Nucl. Phys. A {\bf 944}, 62 (2015).

\bibitem{hamilton_13} J. H. Hamilton, D. Hofmann, and Y. T. Oganessian, 
Annu. Rev. Nucl. Part. Sci. {\bf 63}, 383 (2013).

\bibitem{hong_17} Juhee Jong, G. G. Adamian, and N. V. Antonenko, 
Phys. Lett. B {\bf 764}, 42 (2017).

\bibitem{giuliani_19} S. A. Giuliani, Z. Matheson, W. Nazarewicz, E. Olsen,
P.- G. Reinhard, J. Sadhukhan, B. Schuetrumpf, N. Schunck, and P. Schwerdtfeger, 
Rev. Mod. Phys. {\bf 91}, 011001 (2019).

\bibitem{hofmann_16} S. Hofmann, S. Heinz, R. Mann, J. Maurer, G. M\"unzenberg, 
S. Antalic, W. Barth, H.G. Burkhard, L. Dahl, K. Eberhardt, R. Grzywacz, 
J.H. Hamilton, R.A. Henderson, J.M. Kenneally, B. Kindler, I. Kojouharov, R. Lang, 
B. Lommel, K. Miernik, D. Miller, K.J. Moody, K. Morita, K. Nishio,
A.G. Popeko, J.B. Roberto, J. Runke, K.P. Rykaczewski, S. Saro, C. Scheidenberger, 
H.J. Sch\"ott, D.A. Shaughnessy, M.A. Stoyer, P. Th\"orle-Pospiech, K. Tinschert, 
N. Trautmann, J. Uusitalo, and A.V. Yeremin, Eur. Phys. J. A {\bf 52}, 180 (2016).

\bibitem{myers_66} W. D. Myers and  W. J. Swiatecki, Nuclear Phys. {\bf 81}, 1 (1966).

\bibitem{sobiczewski_66} A. Sobiczewski, F. A. Gareev, and B. N. Kalinkin, 
Phys. Lett. {\bf 22}, 500 (1966).

\bibitem{nilsson_68} S. G. Nilsson,  J. R. Nix, A. Sobiczewski, Z. Szymanski, S. Wycech, 
C. Gustafson, and P. M\"oller,   
Nucl. Phys. A {\bf 115}, 545 (1968).

\bibitem{patyk_91} Z. Patyk and A. Sobiczewski, Nucl. Phys. A {\bf 533}, 132 (1991).

\bibitem{moller_94} P. M\"oller and J. R. Nix, 
J. Phys. G: Nucl. Part. Phys.  {\bf 20}, 1681 (1994).

\bibitem{smolanczuk_95} R. Smolanczuk, J. Skalski, and A. Sobiczewski,  
Phys. Rev. C {\bf 52}, 1871 (1995).

\bibitem{rutz_97} K. Rutz, M. Bender, T. B\"urvenich, T. Schilling, P.-G. Reinhard, 
J. A. Maruhn,  and W. Greiner, Phys. Rev. C {\bf 56}, 238 (1997).

\bibitem{kruppa_00} A. T. Kruppa, M. Bender, W. Nazarewicz, P.-G. Reinhard, T. Vertse, 
and S. \'Cwiok, Phys. Rev. C {\bf 61}, 034313 (2000).

\bibitem{bender_01} M. Bender, W. Nazarewicz , and P.-G. Reinhard, 
Phys. Lett. B {\bf 515}, 42 (2001).

\bibitem{bender_03} M. Bender, P.-H. Heenen, and P.-G. Reinhard, 
Rev. Mod. Phys. \textbf{75}, 121 (2003).

\bibitem{meng_06} J. Meng, H. Toki, S. G. Zhou, S. Q. Zhang, W. H. Long, and L. S. Geng,
Prog. Part. Nucl. Phys. \textbf{57}, 470 (2006).

\bibitem{dobaczewski_15} J. Dobaczewski, A. V. Afanasjev, M. Bender, L. M. Robledo, 
and Yue Shi, Nucl. Phys. A {\bf 944}, 388 (2015).

\bibitem{kuzmina_12}  A. N. Kuzmina, G. G. Adamian, N. V. Antonenko, and W. Scheid, 
Phys. Rev. C {\bf 85}, 014319 (2012).

\bibitem{karpov_12} A. V. Karpov, V. I. Zagrebaev, Y. Martinez Palazuela, L. Felipe Ruiz,
and Walter Greiner, Int. J. Mod. Phys. E {\bf 21}, 1250013 (2012).

\bibitem{zagrebaev_12} V. I. Zagrebaev, A. V. Karpov, and Walter Greiner, 
Phys. Rev. C {\bf 85}, 014608 (2012).

\bibitem{heenen_15} P.-H. Heenen,  J. Skalski, A. Staszczakc, and D. Vretenar, 
Nucl. Phys. A {\bf 944}, 415 (2015).

\bibitem{moller_97}  P. M\"oller, J. R. Nix, and K.-L. Kratz, 
At. Data Nucl. Data Tables {\bf 66}, 131 (1997).

\bibitem{FRDM} P. M\"oller, J. R. Nix, W. D. Myers, and W. J. Swiatecki, 
At. Data Nucl. Data Tables {\bf 59}, 185 (1995).

\bibitem{fiset_72} E. O. Fiset and J. R. Nix, Nucl. Phys. A {\bf 193}, 647 (1972),

\bibitem{sarri1}  P. Sarriguren, E. Moya de Guerra, A. Escuderos, and A. C. Carrizo, 
Nucl. Phys. A {\bf 635}, 55 (1998).

\bibitem{sarri_99} P. Sarriguren, E. Moya de Guerra, and A. Escuderos, 
Nucl. Phys. A {\bf 658}, 13 (1999).

\bibitem{sarri2} P. Sarriguren, E. Moya de Guerra, and A. Escuderos, 
Nucl. Phys. A {\bf 691},  631 (2001).

\bibitem{sarri_odd} P. Sarriguren, E. Moya de Guerra, and A. Escuderos, 
Phys. Rev. C {\bf 64},  064306 (2001).

\bibitem{chabanat} E. Chabanat, P. Bonche, P. Haensel, J. Meyer, and R. Schaeffer, 
Nucl. Phys. A {\bf 635}, 231 (1998).

\bibitem{bender_08} M. Bender, G. F. Bertsch, and P.-H. Heenen, 
Phys. Rev. C {\bf 78}, 054312 (2008).

\bibitem{stoitsov_03} M. V. Stoitsov, J. Dobaczewski, W. Nazarewicz, S. Pittel, 
and D. J. Dean, Phys. Rev. C {\bf 68}, 054312 (2003).

\bibitem{vautherin} D. Vautherin and D. M. Brink, Phys. Rev. C {\bf 5}, 626 (1972); 
D. Vautherin, Phys. Rev. C {\bf 7}, 296 (1973).

\bibitem{stoitsov} M. V. Stoitsov, W. Nazarewicz, and S. Pittel Phys. Rev. C {\bf 58}, 
2092 (1998); M. V. Stoitsov, J. Dobaczewski, P. Ring, and S. Pittel, Phys. Rev. 
C {\bf 61}, 034311 (2000).

\bibitem{moller1} J. Krumlinde and P. M\"oller, Nucl. Phys. A {\bf 417}, 419 (1984).

\bibitem{homma} H. Homma, E. Bender, M. Hirsch, K. Muto, H. V. Klapdor-Kleingrothaus, 
and T. Oda, Phys. Rev. C {\bf 54}, 2972 (1996).

\bibitem{sarri_pb3} J. M. Boillos and P. Sarriguren, Phys. Rev. C {\bf 91}, 034311 (2015).

\bibitem{hir2} K. Muto, E. Bender, T. Oda, and H. V. Klapdor-Kleingrothaus, 
Z. Phys. A  {\bf 341}, 407 (1992).

\bibitem{bm} A. Bohr and B. Mottelson, {\em Nuclear Structure}, Vols. I and II, 
(Benjamin, New York 1975).

\bibitem{schunck_10} N. Schunck, J. Dobaczewski, J. McDonnell, J. Mor\'e, W. Nazarewicz, 
J. Sarich, and M. V. Stoitsov, Phys. Rev. C {\bf 81}, 024316 (2010).

\bibitem{robledo_08} S. Perez-Martin and L. M. Robledo, 
Phys. Rev. C {\bf 78}, 014304 (2008).

\bibitem{sarri_wp} P. Sarriguren, R. Alvarez-Rodr\'{\i}guez, and E. Moya de Guerra, 
Eur. Phys. J. A {\bf 24}, 193 (2005).

\bibitem{sarri_rp} P. Sarriguren, Phys. Rev. C {\bf 79}, 044315 (2009); 
Phys. Lett. B {\bf 680}, 438 (2009); 
Phys. Rev. C {\bf 83}, 025801 (2011).

\bibitem{sarri_pb1} P. Sarriguren, O. Moreno, R. Alvarez-Rodr\'{\i}guez, 
and E. Moya de Guerra, Phys. Rev. C {\bf 72}, 054317 (2005).

\bibitem{sarri_pb2} O. Moreno, P. Sarriguren, R. Alvarez-Rodr\'{\i}guez, 
and E. Moya de Guerra, Phys. Rev. C {\bf 73}, 054302 (2006).

\bibitem{sarripere1} P. Sarriguren and J. Pereira, Phys. Rev. C {\bf 81}, 064314 (2010).

\bibitem{sarripere2} P. Sarriguren, A. Algora, and J. Pereira, 
Phys. Rev. C {\bf 89}, 034311 (2014).

\bibitem{sarripere3} P. Sarriguren, Phys. Rev. C {\bf 91}, 044304 (2015).

\bibitem{kiss} P. Sarriguren, A. Algora, and G. Kiss,  
Phys. Rev. C {\bf 98}, 024311 (2018).

\bibitem{sarri_rare} P. Sarriguren, Phys. Rev. C {\bf 95}, 014304 (2017).

\bibitem {sarri_fp1} P. Sarriguren, E. Moya de Guerra, and R. Alvarez-Rodr\'{\i}guez, 
Nucl. Phys. A {\bf 716},  230 (2003).

\bibitem{sarri_fp2} P. Sarriguren, Phys. Rev. C {\bf 87}, 045801 (2013).

\bibitem{sarri_fp3} P. Sarriguren, Phys. Rev. C {\bf 93}, 054309 (2016).

\bibitem{expnacher} E. N\'acher, A. Algora, B. Rubio, J. L. Ta\'\i n, D. Cano-Ott,
S. Courtin, Ph. Dessagne, F. Mar\'echal, Ch. Mieh\'e, E. Poirier,
M. J. G. Borge, D. Escrig, A. Jungclaus, P. Sarriguren, O. Tengblad,
W. Gelletly, L. M. Fraile and G. Le Scornet, Phys. Rev. Lett. {\bf 92}, 232501 (2004).

\bibitem{gove} N.B. Gove and M.J. Martin, Nucl. Data Tables {\bf 10}, 205 (1971).

\bibitem{audi_12}  G. Audi, F. G. Kondev, M. Wang, B. Pfeiffer, X. Sun, J. Blachot, 
and M. MacCormick, Chinese Physics C {\bf 36}, 1157 (2012); 
M. Wang. G. Audi, A. H. Wapstra, F. G. Kondev, M. MacCormick, X. Xu, and B. Pfeiffer, 
Chinese Physics C {\bf 36}, 1603 (2012).

\bibitem{nndc} www.nndc.bnl.gov

\bibitem{ETFSI} Y. Aboussir, J. M. Pearson, A. K. Dutta, F. Tondeur, 
At. Data Nucl. Data Tables {\bf 61}, 127 (1995).

\bibitem{DufloZuker} J. Duflo and A. P. Zuker, Phys. Rev. C {\bf 52}, R23 (1995).

\bibitem{HFB21} S. Goriely, N. Chamel, and J. M. Pearson, 
Phys. Rev. C {\bf 82}, 035804 (2010).

\bibitem{mass_sly4} M. V. Stoitsov, J. Dobaczewski, W. Nazarewicz, and P. Ring, 
Comp. Phys. Comm. {\bf 167}, 43 (2005).

\bibitem{www_masses} www.nuclearmasses.org

\bibitem{gogny} S. Hilaire and M. Girod, Eur. Phys. J. A {\bf 33}, 237 (2007); \\
www-phynu.cea.fr/science\_en\_ligne/carte\_potentiels\_mi\-croscopiques/carte\_potentiel\_nucleaire\_eng.htm

\bibitem{moller_12} P. M\"oller, A. J. Sierk, T. Ichikawa, and H. Sagawa, 
At. Data Nucl. Data Tables {\bf 109-110}, 1 (2016).

\bibitem{parkhomenko_05} A. Parkhomenko and A. Sobiczewski, 
Acta Physica Polonica B {\bf 36}, 3095 (2005).

\bibitem{royer_00} G. Royer, J. Phys. G: Nucl. Part. Phys.  {\bf 26}, 1149 (2000).

\bibitem{viola_66} V. E. Viola, Jr., and G. T. Seaborg, 
J. Inorg. Nucl. Chem. {\bf 28}, 741 (1966).

\bibitem{sobiczewski_89} A. Sobiczewski, Z. Patyk, and S. \'Cwiok, 
Phys. Lett. B {\bf 224}, 1 (1989).

\end{thebibliography}
\end{document}